\documentclass[twocolumn,preprintnumbers,superscriptaddress,nofootinbib]{revtex4}
\usepackage[colorlinks]{hyperref}
\usepackage{multirow}
\usepackage{amssymb,amsmath,graphicx,subfigure,color,rotating,multirow}

\begin{document}
  \title{Study on the Cabibbo-favored ${\overline B}_{d,s}$ ${\to}$ $D_{d,s}^{(*)+}S^{-}$ weak decays in QCD factorization}
  \author{Lili Chen}
  \affiliation{Centre for Theoretical Physics,
              Henan Normal University, Xinxiang 453007, China}

  \author{Chenyang Jing}
  \affiliation{Centre for Theoretical Physics,
              Henan Normal University, Xinxiang 453007, China}

  \author{Kaiyuan Gao}
  \affiliation{Centre for Theoretical Physics,
              Henan Normal University, Xinxiang 453007, China}

  \author{Shuai Xu}
  \affiliation{School of Physics and Telecommunications Engineering,
              Zhoukou Normal University, Zhoukou 466001,
              China}

  \author{Mengfei Zhao}
  \email{zhaomf1994@163.com(Corresponding author)}
  \affiliation{School of Physics and Electrical Engineering,
              Anyang Normal University, Anyang 455000,
              China}


  \begin{abstract}
    Motivated by recent experimental progress and theoretical developments, we investigate the Cabibbo-favored $b\to c$ governed ${\overline B}_{d,s}$ ${\to}$ $D_{d,s}^{(*)+}S^{-}$~($S$$=$$K_0^*(1430)$, $a_0(1450)$) weak decays by considering the next-to-leading (NLO) contributions within QCDF framework. With the updated values of $B_{(s)}\to D_{(s)}^{(*)}$ transition form factors obtained from a covariant light-front quark model, branching ratios are estimated in two scenarios for scalar mesons. It is found that the branching ratios for ${\overline B}^0{\to}D^{+}{a_0^-}$ and ${\overline B}_{s}^0{\to}D_{s}^{+}{a_0^-}$ decays can reach up to the order of ${\cal O}(10^{-4})$ in scenario-2 by assuming that the scalar mesons are lowest-lying p-wave states, which deserve high-priority experimental searches and may be observed in the ongoing LHCb and SuperKEKB experiments.
  \end{abstract}
  \maketitle
\section{Introduction}
\label{sec01}
  The scalar meson spectrum and internal structure remains an interesting and somewhat contentious topic in hadronic physics. Two primary scenarios have been proposed to explain the internal structure of scalar mesons. In scenario-1~(S1), the scalar mesons with mass below $1\, {\rm GeV}$ are interpreted as the lowest lying $q\overline{q}$ states with a unit of orbital angular momentum, forming an SU(3) nonet that includes particles such as $a_0(980)$, $f_0(980)$ and $K_0^*(700)$. And the ones with mass above $1\, {\rm GeV}$ are treated as the first orbitally excited $q\overline{q}$ states, forming another SU(3) nonet, such as $a_0(1450)$, $f_0(1370)$, $K_0^*(1430)$, etc. In comparison, scenario-2~(S2), proposed by Jaffe~\cite{Jaffe:1976ig,Jaffe:1976ih}, suggests that the light scalar mesons are predominantly tetraquark states ($qq\overline{q}\overline{q}$), while mass above $1~{\rm GeV}$ scalar mesons are considered the lowest lying $q\overline{q}$ p-wave states. The S2 supported by some lattice calculations~\cite{Mathur:2006bs,Prelovsek:2010kg} and mesonic spectroscopy data~\cite{Close:2002zu} has widely been accepted, and is compelling because a scalar meson with a unit orbital angular momentum should naturally have a higher mass above $1\, {\rm GeV}$ in the $q\overline{q}$ model. The two scenarios offer contrasting interpretations of scalar mesons, particularly in the context of their mass and structure.

  $B$ mesons exhibit rich decay patterns, providing a prolific source of light scalar mesons, such as $K_0^*(1430)$, $a_0(1450)$ and $f_0(980)$. Consequently, hadronic decays involving scalar mesons provide a valuable laboratory for investigating the properties and possible internal structures of scalar states. With the continued improvement of $B$ factories, increasing efforts have been devoted to measuring $B_{u,d,s}$$\to$$SM$($S$ denotes the scalar meson) decay modes by Belle~\cite{Belle:2005rpz,Belle:2004khm,Belle:2013vat,Belle:2010uya}, BABAR~\cite{BaBar:2012iuj,BaBar:2011vfx,BaBar:2009mcf,BaBar:2008lpx,BaBar:2008ozy,BaBar:2008lan,BaBar:2011ryf,BaBar:2007eog} and LHCb~\cite{LHCb:2019vww,LHCb:2019bnl} collaborations, such as the $B$ ${\to}$
  $K^{\ast}_{0}(1430)^{+}{\pi}^{-}$,
  $K^{\ast}_{0}(1430)^{+}{\omega}$,
  $K^{\ast}_{0}(1430)^{0}\overline{K}^{\ast}_{0}(1430)^{0}$,
  $K^{\ast}_{0}(1430)^{0}{\pi}^{+}{\gamma}$ decays
  have been measured by Belle, BaBar and LHCb groups \cite{ParticleDataGroup:2024cfk}. In addition, the $b\to c$ induced decays $B^{0}$$\to$${\overline D}^{0}$$f_0(980)$, $B^{0}$$\to$${\overline D}^{0}$$f_0(500)$ were first reported by the LHCb collaboration~\cite{LHCb:2015klp} and the first evidence of $B^{0}$$\to$${\overline D}^{0}$$f_0(2020)$ are presented, in which the branching ratios can reach up to ${\cal O}(10^{-4})$. The ongoing Belle II and LHCb experiments, as well as the forthcoming experiments such as CEPC, FCC-ee and HL-LHC, are expected to provide huge amount of data on $B$ meson weak decay, thereby providing excellent opportunities to measure the $B$ weak decays at $10^{-4}$ magnitude and advancing our understanding of the structure of the scalar states.

  From a theoretical aspect, many charmless hadronic $B$ decays involving scalar meson final state have been studied in detail with QCD-inspired theoretical approaches. For example, the generalized factorization approach~\cite{Cheng:2020ipp}, QCD factorization (QCDF)~\cite{Cheng:2016shb,Cheng:2014uga,Cheng:2013dua,Cheng:2007si,
  El-Bennich:2009gqk,Li:2015zra,Cheng:2005nb,Cheng:2005ye,Cheng:2007st,Cheng:2010sn,
  Cheng:2013fba,Qi:2018syl,Chen:2021oul,Chen:2023pms},  perturbative QCD approach (PQCD)~\cite{Wang:2018xux,Zou:2020fax,Zou:2020atb,Zou:2017iau,Li:2019jlp,Wang:2014ira,Rui:2017hks,Rui:2018mxc,Liu:2009xm,Liu:2013lka,
  Liu:2013cvx,Li:2008tk,Zhang:2013efa,Zhang:2012zze,Zhang:2010kw,Zhang:2010af,Liu:2021gnp} and other methods~\cite{El-Bennich:2006rcn,Cheng:2003sm,Kang:2018jzg,Issadykov:2015iba,Cheng:2019tgh,
  Li:2013aca,Li:2015rea,Ghahramany:2009zz,Aliev:2007rq,Yang:2005bv,Han:2013zg,Wang:2015mxa}. To understand the structure of scalar states, one could deduce which scenario is favorable by comparing the experimental data with the theoretical predictions. Most factorization approaches tend to favor S1. In fact, even in the two-quark picture, the quark component is still unclear, such as the observation of $D_s$$\to$$f_0(980)$$\pi^{+}$ decay introduced the probability of the $s{\overline s}$ component of $f_0(980)$, while $\Gamma(J/\psi\to f_0(980)\omega)$$\sim$$\Gamma(J/\psi\to f_0(980)\phi)$ indicated the existence of the non-strange components ~\cite{Cheng:2002ai,Wang:2006ria}. So it is necessary to make more efforts in $B$ meson weak decays involving scalar meson final state. It is noteworthy that the two-body charmed $B\to DS,D^*S$ decays have been analyzed by employing the perturbative QCD~(PQCD) factorization approach~\cite{Zou:2016yhb,Zou:2017iau}. The study of scalar states in $b\to c$ induced $B_{(s)}$ meson weak decays is particularly advantageous due to the larger branching fractions compared to $B$  decays to light meson final states. The branching fractions of the CKM-favored $B$ decay modes are in the range of ${\cal O}(10^{-4})$$\sim$${\cal O}(10^{-7})$~\cite{Zou:2017iau}, which can be tested in the LHCb experiment and by Belle-II in the near future. The $B^{+}$$\to$${\overline D}^{(*)0}a_0^+(980/1450)$ and $B^{+}$$\to$${\overline D}^{(*)-}a_0^+(980/1450)$ decays have large branching ratios, reaching up to ${\cal O}(10^{-4})$~\cite{Zou:2017iau}, which are measurable with the on-going and forthcoming experiments. However the large uncertainties induced by the nonperturbative parameters might lower the predictive power and blur the distinction between the two scenarios even in favorable cases. In this work, we perform a detailed study of the charmed hadronic $B$ meson weak decays $\overline B$$\to$$DS, D^*S$ within QCDF framework, as to deepen our understanding of scalar meson properties and provide valuable theoretical references for ongoing and future experimental studies.

  As is well known, theoretical predictions of the QCD factorization formula for $B$ meson two-body decays to a heavy-light final state induced by the $b$$\to$$c$ transition are numerically more precise and reliable than those for light-light final state decays. Within the QCDF framework, $B\to D^{(*)}S$ decays are theoretically clean and QCDF approach is expected to work well for them, where the emission particle is scalar meson and recoil meson picks up the heavy spectator $c$ quark and forming heavy meson $D^{(*)}$.
  At leading power in $\Lambda_{\rm QCD}/m_b$, these decays receive contributions from neither the penguin operators nor the penguin topology which are dominated by the color-allowed tree topology that receives only vertex corrections. Annihilation contributions and spectator scattering effects are power suppressed in heavy-light final states containing a heavy meson~\cite{Beneke:2000ry}. Moreover, the $b$$\to$$c$ induced charmed $B$ meson nonleptonic decays are CKM-favored and exhibit relatively large branching ratios, making them widely studied within the QCDF framework, as detailed in Refs.~\cite{Bauer:1986bm,Deandrea:1993ma,Bauer:2001cu,Li:2008ts,Azizi:2008ty,
  Chen:2011ut,Fu:2011zzo,Faustov:2012mt,Zhou:2015jba,Huber:2015bva,Huber:2016xod}. The two-body hadronic $B$ meson charmed decays are much more complicated than the charmless cases because of the un-negligible $c$ quark mass. Even the calculation becomes significantly more complicated due to the un-negligible $c$ quark mass, certain decay channels, such as ${\overline B}^0$${\to}$ $D^{(*)+}$$K_0^{*-}(1430)$ and ${\overline B}_s^0$${\to}$ $D_s^{(*)+}a_0(1450)^{-}$, are free of the weak annihilation contribution, making them theoretically clean and particularly suitable for analysis within the QCDF framework. Motivated by these considerations, we perform a detailed study of the exclusive $B$ meson weak decays into a heavy meson ($D$ or $D^*$) and a scalar meson($K_0^*(1430)$ or $a_0(1450)$), ${\overline B}_{d,s}$ ${\to}$ $D_{d,s}^{(*)+}S^{-}$~($S$$=$$K_0^*(1430)$, $a_0(1450)$) based on the two scenarios.

  This paper is organized as follows. In Sect.\ref{sec02}, we will present the theoretical framework and the ${\overline B}_{d,s}$ ${\to}$ $D_{d,s}^{(*)+}S^{-}$ decay amplitudes with QCDF approach. The numerical results and phenomenological discussions will be presented in Sect.\ref{sec03}. The last section is our conclusion.
  \section{theoretical framework and calculation}
  \label{sec02}
The $B$ meson two-body $\overline{B}_{d,s} \to D_{d,s}^{(*)+}S^{-}$ ($S=K_0^*(1430), a_0(1450)$) decays receive contributions from neither the penguin operators nor the penguin topology, which are dominated by the color-allowed tree topology.
In the Standard Model (SM), the low energy effective Hamiltonian responsible for the $\overline{B}_{d,s} \to D_{d,s}^{(*)+}S^{-}$ decays induced by the $b \to c$ transition is given as
  \begin{equation}
  {\cal H}_{\rm eff}\ =\ \frac{G_{F}}{\sqrt{2}}\,
   \sum\limits_{q=d,s}\, V_{cb} V_{uq}^{\ast}\,
   \Big\{ C_{1}({\mu})\,Q_{1}^{c}({\mu})
         +C_{2}({\mu})\,Q_{2}^{c}({\mu}) \Big\}
   + {\rm h.c.}
  \label{hamilton},
   \end{equation}
  with the relevant effective local tree four-quark operators $Q_{1,2}^{c}(\mu)$ are defined as follows,
  \begin{eqnarray}
    Q_{1}^{c} &=&
  [ \overline{c}_{\alpha}{\gamma}_{\mu}(1-{\gamma}_{5})b_{\alpha} ]
  [ \overline{q}_{\beta} {\gamma}^{\mu}(1-{\gamma}_{5})u_{\beta}  ]
    \label{q1}, \\
    Q_{2}^{c} &=&
  [ \overline{c}_{\alpha}{\gamma}_{\mu}(1-{\gamma}_{5})b_{\beta} ]
  [ \overline{q}_{\beta}{\gamma}^{\mu}(1-{\gamma}_{5})u_{\alpha} ]
    \label{q2},
  \end{eqnarray}
  where ${\alpha}$ and ${\beta}$ are color indices and the
  sum over repeated indices is understood.
  $G_{F}$ is the Fermi coupling constant, the Wilson coefficients $C_{1,2}(\mu)$,  and $V_{cb} V_{uq}^{\ast}$~($q=d, s$) are products of the Cabibbo-Kobayashi-Maskawa~(CKM) matrix elements. Using the Wolfenstein parameterization, the CKM factors can be expanded as a power series in the small Wolfenstein parameter $\lambda$,
  \begin{eqnarray}
  V_{cb}V_{ud}^{\ast} &=&
               A{\lambda}^{2}
  - \frac{1}{2}A{\lambda}^{4}
  - \frac{1}{8}A{\lambda}^{6}
  +{\cal O}({\lambda}^{8})
  \label{eq:ckm01}, \\
  V_{cb}V_{us}^{\ast} &=& A{\lambda}^{3}
  +{\cal O}({\lambda}^{8})
  \label{eq:ckm02}\,.
  \end{eqnarray}
  To obtain the decay amplitude, the remaining and the
  most intricate works are to calculate accurately
  hadronic matrix elements of local operators, $\langle M_1 M_2|O_i| B  \rangle $. In the QCDF, the hadronic matrix element of each operator can be written as the convolution integrals of the scattering kernel with the distribution amplitudes~(DAs) of the participating mesons~\cite{Beneke:1999br,Beneke:2001ev,Beneke:2003zv}.
  According to the QCDF's power counting rules, at the leading power in $\Lambda_{\rm QCD}/m_b$, ${\overline B}_{d,s}$ ${\to}$ $D_{d,s}^{(*)+}S^{-}$~($S$$=$$K_0^*(1430)$, $a_0(1450)$) nonleptonic decays do not receive contributions from the penguin topology which are dominated by the color-allowed tree topology that receives only vertex corrections. As well as, interactions with the annihilation diagrams and the spectator degrees of freedom in the $B$ meson are power-suppressed in the heavy-light final states~\cite{Beneke:2000ry}, making them theoretically clean and particularly suitable for analysis within the QCDF framework.
  With the QCDF master formula, hadronic matrix elements
  could be written as \cite{Beneke:2000ry}:
   \begin{equation}
  {\langle}D^{(*)}S{\vert}Q_{i}{\vert}{\overline B}{\rangle} =
   \sum\limits_{i} F_{i}^{ B{\to}D }
  {\int}\,dx\, H_{i}^{I}(x)\,{\Phi}_{S}(x)
   \label{hadronic},
   \end{equation}
  where transition form factor $F_{i}^{ B{\to}D }$
  and light cone distribution amplitudes ${\Phi}_{S}(x)$ of the
  emitted scalar meson are nonperturbative input parameters,
  hard scattering kernels $H_{i}^{I}(x)$ are computable order by
  order with the perturbation theory in principle. The leading twist two-valence-particle distribution
  amplitudes (DAs) of scalar meson are defined in terms of Gegenbauer
  polynomials~\cite{Cheng:2005ye,Cheng:2005nb}:
   \begin{equation}
  {\Phi}_{S}(x)={\overline f}_S\,6\,x\overline{x}
   \sum\limits_{n=0}^{\infty}
   b_{n}^{S}\, C_{n}^{3/2}(x-\overline{x})
   \label{twist},
   \end{equation}
  where $x$ denotes the momentum fractions and $\overline{x}$ $=$ $1$ $-$ $x$. The scale-dependent scalar decay constant for emitted scalar meson ${\overline f}_S$ and the Gegenbauer moments $b_{n}^{S}$, corresponding to the expansion coefficients of Gegenbauer polynomials $C_{n}^{3/2}(x-\overline{x})$, are hadronic parameters.
\begin{figure*}
  \centering
  \subfigure[]{\includegraphics[width=0.23\textwidth]{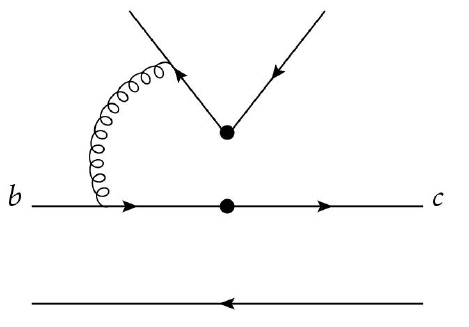}}\hfill
  \subfigure[]{\includegraphics[width=0.23\textwidth]{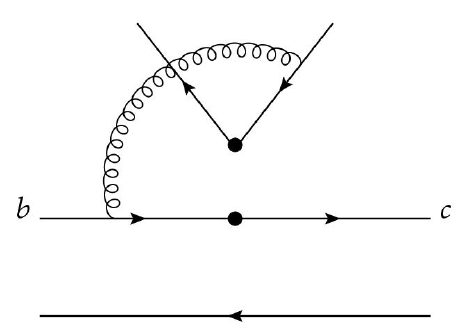}}\hfill
  \subfigure[]{\includegraphics[width=0.23\textwidth]{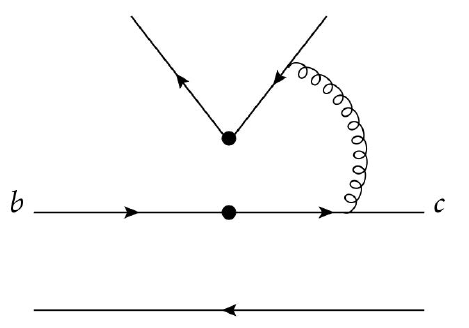}}\hfill
  \subfigure[]{\includegraphics[width=0.23\textwidth]{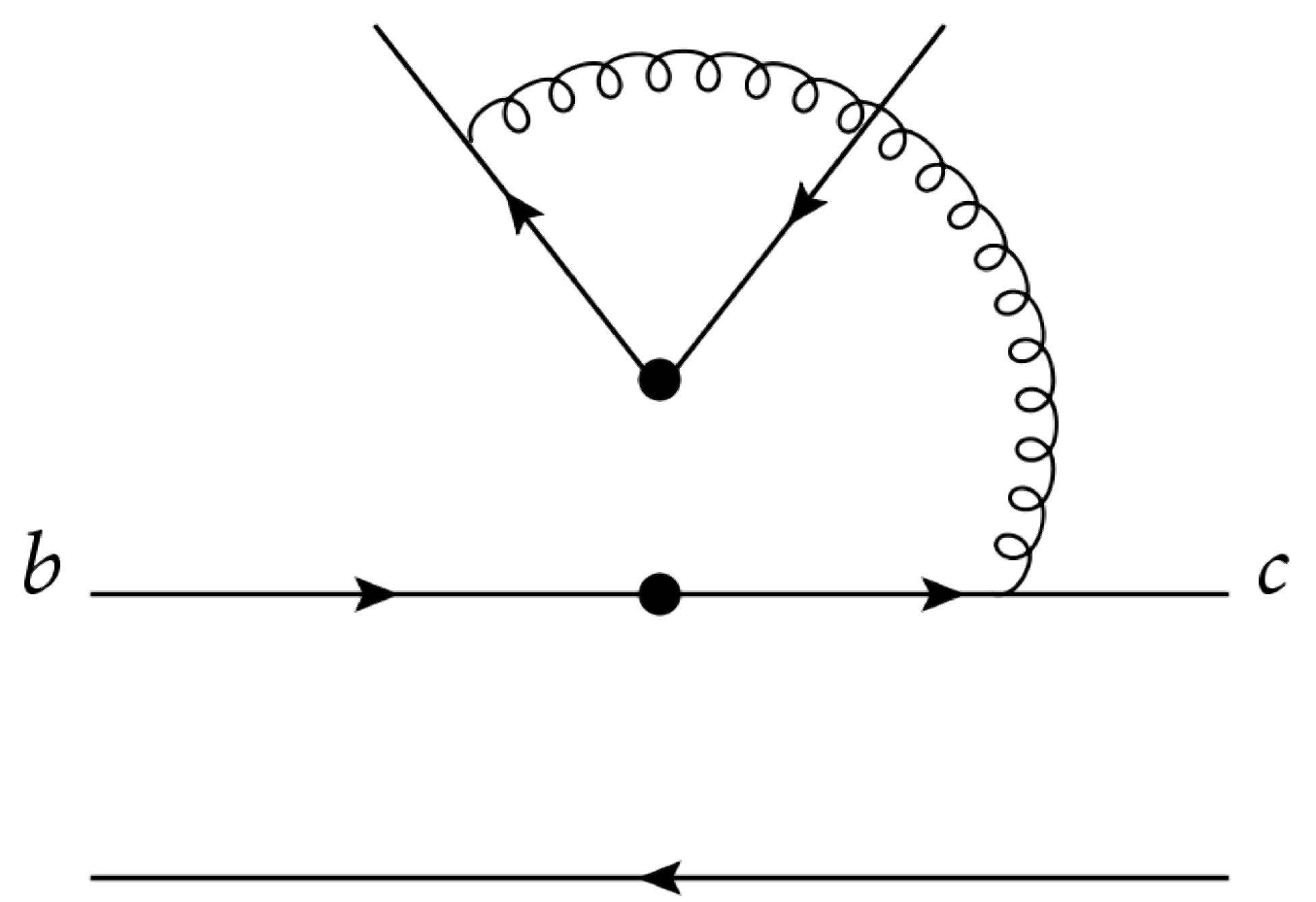}}
  \caption{Next-to-leading-order QCD vertex corrections to the decays ${\overline B}_{d,s}$ ${\to}$ $D_{d,s}^{(*)+}S^{-}(S=K_{0}^{\ast}(1430), a_{0}(1450))$ in QCDF.}
  \label{vc1}
\end{figure*}

  According to the QCDF master~Eq.(\ref{hadronic}), the decay amplitudes for ${\overline B}_{d,s}$ ${\to}$ $D_{d,s}^{(*)+}S^{-}(S=K_{0}^{\ast}(1430),a_{0}(1450))$ weak decay could be written as
\begin{widetext}
   \begin{eqnarray}
  {\cal A}({\overline B}{\to}D^{+}S^{-}) &=&
     \frac{G_{F}}{\sqrt{2}}\, V_{cb} V_{uq}^{\ast}\, a_{1}\,
  {\overline f}_S(m_B^2-m_D^2)F_0^{B\to D}(m_S^2)
   \label{ampDS}\,,\\
   {\cal A}({\overline B}{\to}D^{*+}S^{-}) &=&
     -\frac{G_{F}}{\sqrt{2}}\, V_{cb} V_{uq}^{\ast}\, a_{1}\,
  {\overline f}_S\,2m_B\,p_c\,A_0^{B\to D^*}(m_S^2)
   \label{ampDvS}\,.
   \end{eqnarray}
\end{widetext}
  where $F_0^{B\to D}$ and $A_0^{B\to D^*}$ are the $B\to D^{(*)}$ transition form factors. Up to the next-to-leading order (NLO) in the coupling $\alpha_s$, the QCDF coefficient $a_{1}$ including the nonfactorizable contributions from QCD vertex corrections, is expressed as \cite{Cheng:2005ye,Cheng:2005nb}:
  \begin{equation}
   a_{1}
    =( C_{1}^{\rm NLO}+\frac{1}{N_{c}}\,C_{2}^{\rm NLO} )\, {\overline\mu_S}^{-1}
    + \frac{{\alpha}_{s}}{4{\pi}}\, \frac{C_{F}}{N_{c}}\,
      C_{2}^{\rm LO}\, V
   \label{a1}\,,
  \end{equation}
  where the color factor $C_F$$=$$(N_c^2-1)/(2N_c)$ and the color number $N_c$$=$$3$. The definition of the factor $\overline{\mu}_{S}({\mu})$ is
    \begin{equation}
    \overline{\mu}_{S}({\mu}) \, =\,
    \frac{m_{S}}{\overline{m}_{1}({\mu})-\overline{m}_{2}({\mu})}
    \label{chiral-mu},
    \end{equation}
  where $m_{S}$ is the mass of the emission scalar meson,
  and the ${\mu}$-dependent $\overline{m}_{i}$ is the
  $\overline{\rm MS}$ running quark mass and can be
  evaluated with RGE.
  $\overline{m}_{1}$ and $\overline{m}_{2}$
  correspond to the two valence quarks in a
  scalar meson.

  The Feynman diagrams with a scalar meson emitted are presented in Fig.~\ref{vc1}. With the modified minimal subtraction ($\overline{\rm MS}$)
  scheme, the QCD vertex correction factor $V$ is written as \cite{Sun:2007ei}:
  \begin{equation}
  V = 3\,{\log} \Big( \frac{ m_{b}^{2} }{ {\mu}^{2} } \Big)
    + 3\,{\log} \Big( \frac{ m_{c}^{2} }{ {\mu}^{2} } \Big)
    - 18 +{\int}_{0}^{1}dx\, T(x)\,{\Phi}_{S}(x)
  \label{vc01},
  \end{equation}
  where
  \begin{eqnarray}
  T(x) &=&
      \frac{c_{a}}{1-c_{a}}\,{\log}(c_{a})
    - \frac{4\,c_{b}}{1-c_{b}}\,{\log}(c_{b})
      \nonumber \\ &+&
      \frac{c_{d}}{1-c_{d}}\,{\log}(c_{d})
    - \frac{4\,c_{c}}{1-c_{c}}\,{\log}(c_{c})
      \nonumber \\ &-&
      r_{c}\, \Big\{ \frac{c_{a}}{(1-c_{a})^{2}}\,{\log}(c_{a})
    + \frac{1}{1-c_{a}} \Big\}
      \nonumber \\ &-&
      r_{c}^{-1}\,\Big\{ \frac{c_{d}}{(1-c_{d})^{2}}\,{\log}(c_{d})
    + \frac{1}{1-c_{d}} \Big\}
      \nonumber \\ &+&
      f(c_{a})-f(c_{b})-f(c_{c})+f(c_{d})
      \nonumber \\ &+&
      2\, {\log}(r_{c}^{2}) \{ {\log}(c_{a}) -{\log}(c_{b}) \}
  \label{vc02},
  \end{eqnarray}
  \begin{equation}
   f(c)\ =\ 2\,{\rm Li}_{2} \Big( \frac{c-1}{c} \Big)
   -{\log}^{2}(c)-\frac{2\,c}{1-c}{\log}(c),
   \label{ff}
   \end{equation}
  where $r_{c} = m_{c}/m_{b}$ and the relations
  \begin{eqnarray}
  c_{a} &=& x\,(1-r_{c}^{2})
  \label{ca}, \\
  c_{b} &=& \overline{x}\,(1-r_{c}^{2})
  \label{cb}, \\
  c_{c} &=& -c_{a}/r_{c}^{2}
  \label{cc}, \\
  c_{d} &=& -c_{b}/r_{c}^{2}
  \label{cd}.
  \end{eqnarray}

 The two-body hadronic $B$ meson charmed decays are much more complicated than the charmless cases because of the un-negligible $c$ quark mass $m_c$. Charmed weak decays of $B$ meson involve four fundamental scales, not only the weak interaction scale $M_W$, the $b$ quark mass $m_b$, and the ${\rm QCD}$ scale $\Lambda_{\rm QCD}$, as well as the scale of $c$ quark mass $m_c$. The Wilson coefficients $C_{1,2}(\mu)$, which incorporate strong interaction effects above the scales ${\mu}$$\sim$$m_b$, are calculable with the
  perturbation theory and have properly been evaluated to
  the next-to-leading order (NLO) with the renormalization
  group equation (RGE)~\cite{Buchalla:1995vs}.
   The numerical values of the Wilson coefficients $C_{1,2}$ and effective coefficient $a_{1}$ at scales of ${\mu}$ ${\sim}$
  ${\cal O}(m_{b})$ are listed in Table \ref{tab:ci}.
  It is shown that coefficient $a_{1}$ is renormalization scale independent at the order of ${\alpha}_{s}$.
\begin{table*}[ht]
\begin{minipage}[t]{0.18\textwidth}
  \vspace{0pt}
  \vspace*{-10pt}
  \caption{The numerical values of the Wilson coefficients $C_{1,2}$ and $a_{1}$ for the $B \to Da_0$ and $B \to DK_0^*$ decay at different scales, where $m_b = 4.78 \text{ GeV}$.}
  \label{tab:ci}
\end{minipage}\hfill
\begin{minipage}[t]{0.80\textwidth}
  \vspace{0pt}
  \begin{ruledtabular}
   \begin{tabular}{c|cc|cc|cc|cc}
     & \multicolumn{2}{c|}{LO \cite{Buchalla:1995vs}}
     & \multicolumn{2}{c|}{NLO \cite{Buchalla:1995vs}}
     & \multicolumn{2}{c|}{QCDF(S1)}
     & \multicolumn{2}{c}{QCDF(S2)} \\ \cline{2-9}

     $\mu$ & $C_{1}$ & $C_{2}$ & $C_{1}$ & $C_{2}$
           & $a_{1}(Da_0)$ & $a_{1}(DK_0^*)$ & $a_{1}(Da_0)$ & $a_{1}(DK_0^*)$ \\
     \hline

     $0.5\,m_{b}$ & $1.165$ & $-0.332$ & $1.125$ & $-0.264$
                  & $0.056 + i\, 0.004$ & $0.134 + i\, 0.002$
                  & $0.057 + i\, 0.005$ & $0.133 + i\, 0.001$ \\

     $m_{b}$      & $1.109$ & $-0.235$ & $1.075$ & $-0.171$
                  & $0.048 + i\, 0.002$ & $0.125 + i\, 0.001$
                  & $0.048 + i\, 0.003$ & $0.124 + i\, 0.000$ \\

     $1.5\,m_{b}$ & $1.084$ & $-0.188$ & $1.054$ & $-0.127$
                  & $0.041 + i\, 0.002$ & $0.118 + i\, 0.001$
                  & $0.042 + i\, 0.002$ & $0.117 + i\, 0.000$ \\

     $2.0\,m_{b}$ & $1.069$ & $-0.159$ & $1.041$ & $-0.099$
                  & $0.036 + i\, 0.001$ & $0.113 + i\, 0.001$
                  & $0.037 + i\, 0.002$ & $0.112 + i\, 0.000$ \\
  \end{tabular}
  \end{ruledtabular}
\end{minipage}
\end{table*}
  \section{Numerical results and discussion}
  \label{sec03}
  In this section, we present the input parameters used in our numerical calculations, which are collected in Table \ref{tab:input}. Nonperturbative parameters, such as decay constants, distribution amplitudes and $B\to D$ transition form factor, play a central role in theoretical predictions, as they are significantly influenced by assumptions about the scalar meson structure.
  In this work, the decay constant and Gegenbauer moment of scalar mesons will be calculated under two distinct scenarios (S1 and S2). The $B\to D$ transition form factors derived from the self-consistent  covariant light-front (CLF) approach will then be updated, thereby ensuring consistency in the theoretical framework. If not specified explicitly, we will take their central values as the default inputs. For the well-determined Fermi coupling constant and lifetimes of $B$ mesons, we take their default values given by PDG~\cite{ParticleDataGroup:2024cfk}.
\begin{table*}[ht]
\begin{minipage}[t]{0.25\textwidth}
  \vspace{0pt}
  \vspace*{-10pt}
  \caption{The values of input parameters.}
  \label{tab:input}
\end{minipage}\hfill
\begin{minipage}[t]{0.72\textwidth}
  \vspace{0pt}
  \begin{ruledtabular}
  \begin{tabular}{ll}
    \multicolumn{2}{c}{Wolfenstein parameters} \\
    \hline
    $\lambda = 0.22501 \pm 0.00068$ \cite{ParticleDataGroup:2024cfk}
  & $A = 0.826^{+0.016}_{-0.015}$ \cite{ParticleDataGroup:2024cfk} \\
    \hline

    \multicolumn{2}{c}{masses of scalar mesons and quarks} \\
    \hline
    $m_{a_0(1450)} = 1.439 \pm 0.034~\text{GeV}$ \cite{ParticleDataGroup:2024cfk}
  & $m_{K_0^*(1430)} = 1.425 \pm 0.05~\text{GeV}$ \cite{ParticleDataGroup:2024cfk} \\
    $m_{c} = 1.67 \pm 0.07~\text{GeV}$ \cite{ParticleDataGroup:2024cfk}
  & $m_{b} = 4.78 \pm 0.06~\text{GeV}$ \cite{ParticleDataGroup:2024cfk} \\
    $m_{s}(\mu)/m_{q}(\mu) = 27.3^{+0.7}_{-1.3}$ \cite{ParticleDataGroup:2024cfk}
  & $m_{q} \equiv (m_u+m_d)/2$ \\
    $m_{s}(2~\text{GeV}) = 93^{+11}_{-5}~\text{MeV}$ \cite{ParticleDataGroup:2024cfk}
  & $m_{b}(m_b) = 4.18^{+0.03}_{-0.02}~\text{GeV}$ \cite{ParticleDataGroup:2024cfk} \\
    \hline

    \multicolumn{2}{c}{decay constant and Gegenbauer moments in S1 \cite{Chen:2023pms}} \\
    \hline
    $\overline{f}_{K_{0}^{\ast}(1430)} = 234^{+85}_{-87}~\text{MeV}$
  & $\overline{f}_{a_{0}(1450)} = 256^{+56}_{-54}~\text{MeV}$   \\
    $b_{0}^{K_{0}^{\ast}} = 0.08 \pm 0.01$
  & $b_{1}^{K_{0}^{\ast}} = -0.15 \pm 0.05$ \\
    $b_{2}^{K_{0}^{\ast}} = 0.06 \pm 0.01$
  & $b_{3}^{K_{0}^{\ast}} = -0.09 \pm 0.05$ \\
    $b_{0}^{a_{0}} = 0$
  & $b_{1}^{a_{0}} = -0.17 \pm 0.06$ \\
    $b_{2}^{a_{0}} = 0$
  & $b_{3}^{a_{0}} = -0.19 \pm 0.03$ \\
    \hline

    \multicolumn{2}{c}{decay constant and Gegenbauer moments in S2 \cite{Chen:2023pms}} \\
    \hline
    $\overline{f}_{K_{0}^{\ast}(1430)} = 542^{+180}_{-190}~\text{MeV}$
  & $\overline{f}_{a_{0}(1450)} = 456^{+57}_{-56}~\text{MeV}$  \\
    $b_{0}^{K_{0}^{\ast}} = 0.08 \pm 0.01$
  & $b_{1}^{K_{0}^{\ast}} = -0.13 \pm 0.05$ \\
    $b_{2}^{K_{0}^{\ast}} = -0.03 \pm 0.00$
  & $b_{3}^{K_{0}^{\ast}} = -0.01 \pm 0.00$  \\
    $b_{0}^{a_{0}} = 0$
  & $b_{1}^{a_{0}} = -0.17 \pm 0.03$ \\
    $b_{2}^{a_{0}} = 0$
  & $b_{3}^{a_{0}} = -0.42 \pm 0.22$
  \end{tabular}
  \end{ruledtabular}
\end{minipage}
\end{table*}

  There are two definitions of the decay constants for the scalar mesons
    \begin{equation}
   {\langle} S(p)\,{\vert} \overline{q}_{1}\,{\gamma}^{\mu}\, q_{2}\,
   {\vert}\, 0 {\rangle}\, =\, f_{S}\, p^{\mu}
    \label{decay-constant-fs},
    \end{equation}
    \begin{equation}
   {\langle} S(p)\,{\vert} \overline{q}_{1}\, q_{2}\,
   {\vert}\, 0 {\rangle}\, =\, m_{S}\, \overline{f}_{S}({\mu})
    \label{decay-constant-barfs}.
    \end{equation}
   The scale-dependent scalar decay constant $\overline{f}_{S}({\mu})$
   and the vector decay constant $f_{S}$ are related by the
   equation of motion,
    \begin{equation}
     f_{S}\, =\, \overline{f}_{S}({\mu})\, \overline{\mu}_{S}^{-1}({\mu})
    \label{decay-constant-fs-barfs}.
    \end{equation}
   It has been discussed in detail in~\cite{Chen:2023pms}, the vector decay constant $f_{S}$ $=$ $0$ for the $a_{0}^{0}$ meson and the preferable solution scheme is to use the scalar decay constants $\overline{f}_{S}$. This is one of the main reasons for the factor $\overline{\mu}_{S}({\mu})$ in Eq.~(\ref{a1}). As well as the DAs can be expressed with $\overline f_s$ and $b_n^S$ nor $f_s$ and $a_n^S$ in~Eq.(\ref{twist}). We have considered two different scenarios~(S1 and S2) for the scalar meson above 1 GeV within the self-consistent CLF approach in Ref.~\cite{Chen:2023pms}. Here the decay constants $\overline f_S$ and the corresponding Gegenbauer moments $b_n^S$ for $K_{0}^{\ast}(1430)$ and $a_{0}(1450)$ which are summarized in the Table~\ref{tab:input}.

\begin{figure*}
  \centering
  \includegraphics[width=13.5cm]{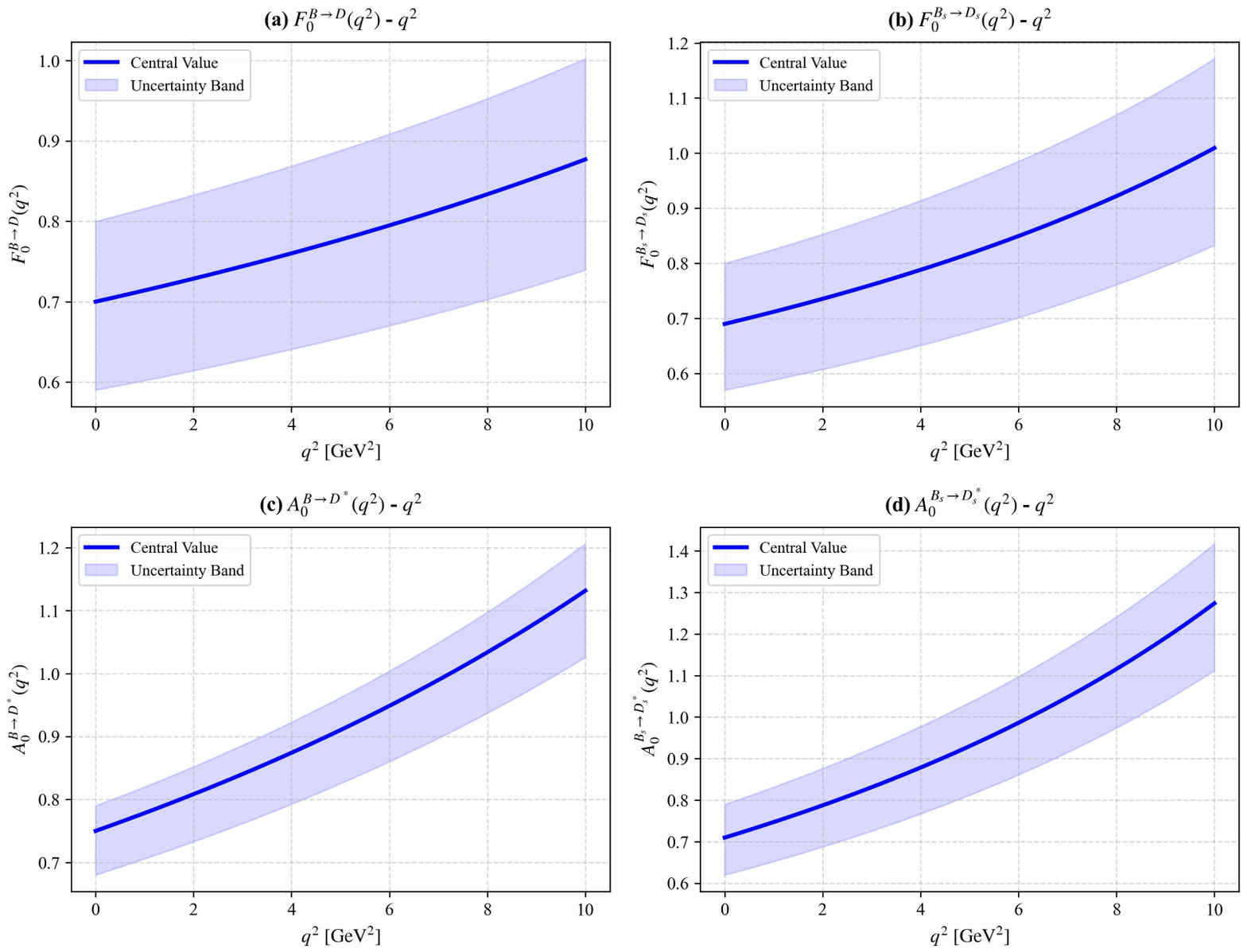}
  \caption{The transition form factors $F_0^{B \rightarrow D}(q^2)$, $F_0^{B_s \rightarrow D_s}(q^2)$, $A_0^{B \rightarrow D^*}(q^2)$, and $A_0^{B_s \rightarrow D_s^*}(q^2)$ as functions of the momentum transfer $q^2$ within the BCL $z$-expansion parameterization approach. The solid blue curves represent the theoretical central values, while the light blue shaded areas signify the standard propagated uncertainties inherited from the non-perturbative input parameters $F(0)$ and $b_1$.}\label{formfactor1}
\end{figure*}

  The transition form factors for $B\to D\,,D^*$ transitions concerned in this paper can be defined as~\cite{Wirbel:1985ji,Beneke:2000wa}:

\begin{widetext}
\begin{eqnarray}\label{FF}
  {\langle}D(p^{\prime\prime})|{\overline c}\gamma_{\mu}b|B(p^\prime){\rangle}
  &=&\bigg(P_\mu-\frac{m_B^2-m_D^2}{q^2}q_\mu\bigg)F_1(q^2)
  +\frac{m_B^2-m_D^2}{q^2}q_\mu F_0(q^2)\,,
  \\
  {\langle}D^{*}(\epsilon,p^{\prime\prime})|{\overline q}\gamma_{\mu}b|B(p^\prime){\rangle}
  &=&\frac{iV(q^2)}{m_B+m_{D^{*}}}\varepsilon_{\mu\nu\alpha\beta} \epsilon^{*\nu}P^{\alpha}q^{\beta}\,,
  \\
  {\langle}D^{*}(\epsilon,p^{\prime\prime})|{\overline q}\gamma_{\mu}\gamma_{5}b|B(p^\prime){\rangle}
  &=&2m_{D^{*}}\frac{\epsilon^*\cdot P}{q^2}q_\mu A_0(q^2)+(m_B+m_{D^{*}})\bigg(\epsilon_\mu^*-\frac{\epsilon^*\cdot P}{q^2}q_\mu\bigg)A_1(q^2)\nonumber \\
  &&-\frac{\epsilon^*\cdot P}{m_B+m_{D^{*}}}\bigg(P_\mu-\frac{m_B^2-m_{D^{*}}^2}{q^2}q_\mu\bigg)A_2(q^2)\,,
\end{eqnarray}
\end{widetext}
  where $P_\mu=(p^\prime+p^{\prime\prime}),\,q_\mu=(p^\prime-p^{\prime\prime})$ and $\varepsilon_{0123}=-1$. With the BCL version of the z-series expansion in~Ref.~\cite{Bourrely:2008za}, the momentum dependence of the form factor can be parameterized as the form
\begin{eqnarray}\label{eq:dipo}
F(q^2)=\frac{F(0)}{1-q^2/m^2_{B^*_c}}\left\{1+\sum^{N}_{k=1}\,b_k\,[z(q^2,t_0)^k-z(0,t_0)^k]\right\}\,,
\end{eqnarray}
where $z(q^2,t_0)=\frac{\sqrt{t_{+}-q^2}-\sqrt{t_{+}-t_0}}{\sqrt{t_{+}-q^2}+\sqrt{t_{+}-t_0}}$, $t_{+}=(m_B+m_{D^{(*)}})^2$ and $t_0=(m_B+m_{D^{(*)}})(\sqrt{m_B}-\sqrt{m_{D^{(*)}}})^2$.
The values of parameters $b$ and $F(0)$ for the transition concerned in this work are summarized in Table~\ref{tab:BDFF}. In particular, the form factors from $B_{(s)}\to D_{(s)}^{*}$ transition have been updated by a self-consistent CLF, compared with the results given in Ref.~\cite{Chang:2019mmh}.
The reason for the difference in the values of the form factors is that the parameters $\beta$ and quark mass have different values.
As illustrated in Fig.~\ref{formfactor1}, the four transition form factors $F_0^{B \rightarrow D}$, $F_0^{B_s \rightarrow D_s}$, $A_0^{B \rightarrow D^*}$, and $A_0^{B_s \rightarrow D_s^*}$ evaluated within the self-consistent CLF approach display a smooth and monotonic dependence across the evaluated range of $q^2 \in [0, 10]\ \mathrm{GeV^2}$. For the Cabibbo-favored $\overline{B}_{d,s}\rightarrow D_{d,s}^{(*)+}S^{-}$ weak decays investigated in this work, the momentum transfer is physically fixed near the mass squared of the emitted scalar mesons ($q^2 = m_S^2 \approx 2.07~\mathrm{GeV^2}$). Within this low-$q^2$ kinematic region, the form factors remain well-behaved and far from the resonance poles, which fundamentally ensures the perturbativity and theoretical cleanness of the hard-scattering kernels within the QCDF framework.
\begin{table}[ht]
\caption{Form factors of $B\to D^{(*)}$ and $B_s\to D_s^{(*)}$ transitions with the self-consistent CLF approach. The errors arise from the parameters $\beta$ and quark masses.}
\label{tab:BDFF}
\begin{ruledtabular}
\begin{tabular}{lcc}
  Form Factors & $F(0)$ & $b_1$ \\
  \hline
  $F_0^{B\to D}$    & $0.70^{+0.10}_{-0.11}$ & $1.20^{+0.03}_{-0.03}$  \\
  $V^{B\to D^*}$    & $0.86^{+0.08}_{-0.09}$ & $-2.32^{+0.17}_{-0.14}$ \\
  $A_0^{B\to D^*}$  & $0.75^{+0.04}_{-0.07}$ & $-2.41^{+0.84}_{-0.12}$ \\
  $A_1^{B\to D^*}$  & $0.72^{+0.07}_{-0.08}$ & $1.41^{+0.09}_{-0.07}$  \\
  $A_2^{B\to D^*}$  & $0.65^{+0.07}_{-0.07}$ & $-1.41^{+0.36}_{-0.38}$ \\
  \hline
  $F_0^{B_s\to D_s}$  & $0.69^{+0.11}_{-0.12}$ & $0.70^{+0.37}_{-0.37}$ \\
  $V^{B_s\to D_s^*}$  & $0.85^{+0.09}_{-0.10}$ & $-3.39^{+0.05}_{-0.10}$ \\
  $A_0^{B_s\to D_s^*}$& $0.71^{+0.08}_{-0.09}$ & $-3.59^{+0.07}_{-0.12}$ \\
  $A_1^{B_s\to D_s^*}$& $0.69^{+0.08}_{-0.09}$ & $0.31^{+0.16}_{-0.23}$ \\
  $A_2^{B_s\to D_s^*}$& $0.63^{+0.08}_{-0.08}$ & $-2.46^{+0.25}_{-0.29}$ \\
\end{tabular}
\end{ruledtabular}
\end{table}
  In the rest frame of $B$ particle, branching ratio
  for nonleptonic ${\overline B}$ ${\to}$ $D^{(*)}S$ weak decays can
  be written as
   \begin{equation}
  {\cal B}r({\overline B}{\to}D^{(*)}S)\ =\ \frac{1}{8{\pi}}\,
   \frac{p_{\rm cm}}{m_B^{2}{\Gamma}_{B}}\,
  {\vert}{\cal A}({\overline B}{\to}D^{(*)}S){\vert}^{2}
   \label{br},
   \end{equation}
 where the momentum of final states is
   \begin{equation}
   p_{\rm cm}\ =\
   \frac{ \sqrt{ [m_B^{2}-(m_{D^{(*)}}+m_{S})^{2}]
                 [m_B^{2}-(m_{D^{(*)}}-m_{S})^{2}] }  }
       { 2\,m_{B} }
   \label{pcm}.
   \end{equation}

 All of these decays only occur through tree operators, the $CP$ asymmetries are null in SM. So we only present the numerical results on the branching ratios
 for the ${\overline B}$ ${\to}$ $D^{(*)}S$ decays in Table~\ref{tab:brS2} and \ref{tab:brS1}. The theoretical uncertainties of the results come from the CKM parameters, the nonperturbative parameters(decay constants and Gegenbauer moments) and transition form factors respectively. The large uncertainties from the nonperturbative parameters might lower our prediction power. Therefore more reliable nonperturbative approaches and experimental data are needed for describing the nature of the scalar mesons in the future. For the sake of comparison, previous results of Ref.~\cite{Zou:2017iau} evaluated with PQCD approach are also listed. The following are some comments:
\begin{table*}[ht]
\begin{minipage}[t]{0.25\textwidth}
  \vspace{0pt}
  \vspace*{-10pt}
  \caption{The branching ratios for the ${\overline B}$ ${\to}$ $D^{(*)}S$ decays in S1.}
  \label{tab:brS1}
\end{minipage}\hfill
\begin{minipage}[t]{0.72\textwidth}
  \vspace{0pt}
  \begin{ruledtabular}
  \begin{tabular}{llll}
    Decay mode & NF$(\times 10^{-6})$ & QCDF$(\times 10^{-5})$ & Ref.~\cite{Zou:2017iau}$(\times 10^{-5})$ \\
    \hline

    ${\cal B}r({\overline B}^0{\to}D^{+}{a_0^-})$
    & $0.08^{+ 0.00+ 0.04+ 0.02}_{- 0.00- 0.03- 0.02}$
    & $3.80^{+ 0.16+ 1.85+ 1.30}_{- 0.14- 1.47- 1.12}$
    & $4.09^{+ 2.63+ 1.59+ 0.23}_{- 2.09- 0.86- 0.10}$
    \\

    ${\cal B}r({\overline B}^0{\to}{D^+}{K_0^{*-}})$
    & $4.77^{+ 0.21+ 4.09+ 1.50}_{- 0.19- 2.89- 1.41}$
    & $1.16^{+ 0.05+ 0.99+ 0.36}_{- 0.05- 0.71- 0.34}$
    & $0.97^{+ 0.45+ 0.14+ 0.06}_{- 0.34- 0.12- 0.03}$
    \\

    ${\cal B}r({\overline B}^0_s{\to}{D_s^{+}}{a_0^-})$
    & $0.07^{+ 0.00+ 0.04+ 0.03}_{- 0.00- 0.03- 0.02}$
    & $3.78^{+ 0.15+ 1.84+ 1.32}_{- 0.14- 1.60- 1.21}$
    & $10.3^{+  5.8+  1.9+  0.5}_{-  4.9-  1.8-  0.3}$
    \\

    ${\cal B}r({\overline B}^0_s{\to}{D_s^{+}}K_0^{*-})$
    & $4.74^{+ 0.20+ 4.07+ 1.66}_{- 0.19- 2.87- 1.52}$
    & $1.15^{+ 0.05+ 0.99+ 0.40}_{- 0.05- 0.70- 0.37}$
    & $0.36^{+ 0.24+ 0.13+ 0.02}_{- 0.18- 0.08- 0.01}$
    \\

    ${\cal B}r({\overline B}^0{\to}D^{*+}{a_0^-})$
    & $0.06^{+ 0.00+ 0.03+ 0.01}_{- 0.00- 0.02- 0.01}$
    & $3.34^{+ 0.14+ 1.62+ 0.33}_{- 0.13- 1.44- 0.54}$
    & $26.8^{+ 12.7+  7.5+  1.5}_{- 11.1-  7.3-  0.6}$
    \\

    ${\cal B}r({\overline B}^0{\to}{D^{*+}}{K_0^{*-}})$
    & $4.22^{+ 0.18+ 3.62+ 0.42}_{- 0.17- 2.55- 0.68}$
    & $1.02^{+ 0.04+ 0.88+ 0.10}_{- 0.04- 0.62- 0.17}$
    & $0.75^{+ 0.26+ 0.32+ 0.05}_{- 0.24- 0.27- 0.02}$
    \\

    ${\cal B}r({\overline B}^0_s{\to}{D_s^{*+}}{a_0^-})$
    & $0.06^{+ 0.00+ 0.03+ 0.01}_{-0.00- 0.02- 0.01}$
    & $3.08^{+ 0.13+ 1.49+ 0.74}_{-0.12- 1.73- 0.73}$
    & $23.0^{+ 11.1+  8.9+  1.2}_{-10.1-  7.8-  0.6}$
    \\

    ${\cal B}r({\overline B}^0_s{\to}{D_s^{*+}}K_0^{*-})$
    & $3.88^{+ 0.17+ 3.33+ 0.94}_{- 0.16- 2.35- 0.92}$
    & $0.94^{+ 0.04+ 0.81+ 0.23}_{- 0.04- 0.61- 0.22}$
    & $1.40^{+ 0.55+ 0.44+ 0.09}_{- 0.50- 0.41- 0.04}$
    \\
  \end{tabular}
  \end{ruledtabular}
\end{minipage}
\end{table*}

\begin{table*}[ht]
\begin{minipage}[t]{0.25\textwidth}
  \vspace{0pt}
  \vspace*{-10pt}
  \caption{The branching ratios for the ${\overline B}$ ${\to}$ $D^{(*)}S$ decays in S2.}
  \label{tab:brS2}
\end{minipage}\hfill
\begin{minipage}[t]{0.72\textwidth}
  \vspace{0pt}
  \begin{ruledtabular}
  \begin{tabular}{llll}
    Decay mode & NF$(\times 10^{-6})$ & QCDF$(\times 10^{-5})$ & Ref. \cite{Zou:2017iau}$(\times 10^{-5})$ \\
    \hline

    ${\cal B}r({\overline B}^0{\to}D^{+}{a_0^-})$
    & $0.24^{+ 0.01+ 0.06+ 0.08}_{- 0.01- 0.06- 0.07}$
    & $11.83^{+ 0.48+ 3.15+ 3.72}_{- 0.45- 2.75- 3.48}$
    & $0.92^{+ 0.57+ 0.41+ 0.05}_{- 0.43- 0.24- 0.02}$
    \\

    ${\cal B}r({\overline B}^0{\to}{D^+}{K_0^{*-}})$
    & $25.61^{+ 1.11+ 19.83+ 8.06}_{- 1.03- 14.81- 7.54}$
    & $6.15^{+ 0.26+ 4.76+ 1.94}_{- 0.25- 3.56- 1.81}$
    & $0.79^{+ 0.46+ 0.14+ 0.05}_{- 0.38- 0.11- 0.03}$
    \\

    ${\cal B}r({\overline B}^0_s{\to}{D_s^{+}}{a_0^-})$
    & $0.24^{+ 0.01+ 0.06+ 0.08}_{- 0.01- 0.06- 0.08}$
    & $11.76^{+ 0.48+ 3.13+ 4.12}_{- 0.45- 2.71- 3.77}$
    & $4.35^{+ 3.28+ 1.21+ 0.23}_{- 2.43- 0.97- 0.11}$
    \\

    ${\cal B}r({\overline B}^0_s{\to}{D_s^{+}}K_0^{*-})$
    & $25.43^{+ 1.11+ 19.70+ 8.92}_{- 1.02- 14.71- 8.15}$
    & $6.11^{+ 0.26+ 4.73+ 2.14}_{- 0.25- 3.53- 1.96}$
    & $0.39^{+ 0.21+ 0.04+ 0.02}_{- 0.16- 0.04- 0.01}$
    \\

    ${\cal B}r({\overline B}^0{\to}D^{*+}{a_0^-})$
    & $0.21^{+ 0.09+ 0.06+ 0.02}_{- 0.08- 0.05- 0.03}$
    & $10.40^{+ 0.42+ 2.76+ 1.03}_{- 0.39- 2.40- 1.68}$
    & $11.3^{+  7.4+  2.1+  0.7}_{-  5.9-  2.7-  0.2}$
    \\

    ${\cal B}r({\overline B}^0{\to}{D^{*+}}{K_0^{*-}})$
    & $22.63^{+ 0.98+ 17.53+ 2.23}_{- 0.91- 13.09- 3.66}$
    & $5.44^{+ 0.23+ 4.21+ 0.54}_{- 0.22- 3.14- 0.88}$
    & $0.09^{+ 0.14+ 0.02+ 0.01}_{- 0.07- 0.02- 0.01}$
    \\

    ${\cal B}r({\overline B}^0_s{\to}{D_s^{*+}}{a_0^-})$
    & $0.19^{+ 0.08+ 0.05+ 0.05}_{- 0.07- 0.04- 0.04}$
    & $9.57^{+ 0.39+ 2.54+ 2.31}_{- 0.36- 2.24- 2.27}$
    & $6.20^{+ 4.58+ 1.46+ 0.33}_{- 3.62- 1.64- 0.16}$
    \\

    ${\cal B}r({\overline B}^0_s{\to}{D_s^{*+}}K_0^{*-})$
    & $20.83^{+ 0.90+ 16.14+ 5.01}_{- 0.84- 12.05- 4.94}$
    & $5.00^{+ 0.21+ 3.88+ 1.21}_{- 0.20- 2.89- 1.19}$
    & $0.24^{+ 0.27+ 0.06+ 0.01}_{- 0.20- 0.08- 0.01}$
    \\
  \end{tabular}
  \end{ruledtabular}
\end{minipage}
\end{table*}
\begin{itemize}
\item From Table~\ref{tab:brS1} and Table~\ref{tab:brS2}, it can be found that the branching ratios obtained based on S2 are generally much larger than the ones based on S1. Moreover, the branching ratios ${\cal B}r({\overline B}_{(s)}^0{\to}D_{(s)}^{(*)+}{a_0^-})$ and ${\cal B}r({\overline B}_{(s)}^0{\to}D_{(s)}^{(*)+}{K_0^{*-}})$ in S2 are almost 3 and 5 times larger than that in S1, respectively, which indicate that the branching ratios are sensitive to the scenarios. The main reason comes from the  relatively larger decay constant of $K_0^{*}(1430)$ and $a_0(1450)$ predicted in S2 than in S1.
\item
In the NF scheme, the LO contributions are heavily suppressed by the factor $\overline{\mu}_S(\mu)$ in Eq.~(\ref{a1}). Consequently, the NLO contributions become critically important, generally yielding significant enhancements for both the S1 and S2. This effect can even alter the LO contributions by up to two orders of magnitude, as seen in the ``NF'' columns for the ${\overline B}_{(s)}^0{\to}D_{(s)}^{(*)+}{a_0^-}$ processes in Table \ref{tab:brS1} and \ref{tab:brS2}. Furthermore, this $\overline{\mu}_S(\mu)$ factor is responsible for a notable hierarchy within the NF scheme itself: the branching ratios for the $D_{(s)}^{(*)}K_0^{*}$ final states are consistently larger than those for the $D_{(s)}^{(*)}a_0$ states. This discrepancy arises from the significant mass difference between the up and strange quarks within $\overline{\mu}_S(\mu)$, which dynamically enhances the ${\overline B}_{(s)}^0{\to}D_{(s)}^{(*)}K_0^{*}$ decay channels.
\item
  In QCDF scheme, there exists a clear hierarchical relation between the vector-scalar $D^*S$ final state and pseudoscalar-scalar $DS$ final state for both S1 and S2, as follows:
\begin{widetext}
  \begin{eqnarray}
  {\cal B}r({\overline B}_{(s)}^0{\to}D_{(s)}^+{a_0^-}) >
  {\cal B}r({\overline B}_{(s)}^0{\to}D_{(s)}^{*+}{a_0^-}) >
  {\cal B}r({\overline B}_{(s)}^0{\to}D_{(s)}^+{K_0^{*-}}) >
  {\cal B}r({\overline B}_{(s)}^0{\to}D_{(s)}^{*+}{K_0^{*-}})\,.
  \end{eqnarray}
\end{widetext}
 This hierarchy arises from three dynamical factors. Firstly, sizable enhancement of NLO contributions to the $D_{(s)}^{(*)}a_0$ final state process. Secondly, the CKM factor $|V_{cb}V_{ud}^*|$ responsible for the $\overline{B}_{(s)}^0 \to D_{(s)}^{(*)+} a_0^-$ decays is larger than the CKM factor $|V_{cb}V_{us}^*|$ responsible for the $\overline{B}_{(s)}^0 \to D_{(s)}^{(*)+} K_0^{*-}$ decays. Thirdly, the decay of the pseudoscalar ${\overline B}$ meson into the vector-scalar $D^*S$ state is suppressed due to the orbital angular momentum compared to the $\overline{B}$ decay into the $DS$ state with the same flavor structure.

\item Compared to the PQCD approach in Ref.~\cite{Zou:2017iau}, our QCDF predictions exhibit distinct theoretical divergences, primarily driven by the treatment of annihilation diagrams. In the PQCD framework, interference effects between emission and annihilation diagrams generally result in smaller branching ratios for S2 than S1 (with the exception of the ${\overline B}^0{\to}D^{+}{K_0^{*-}}$ and ${\overline B}_s^0{\to}D_s^{+}{K_0^{*-}}$ decays). A specific manifestation of this divergence is the relation between $\mathcal{B}r(\overline{B}^{0}\rightarrow D^{+}a_{0}^{-})$ and $\mathcal{B}r(\overline{B}_{s}^{0}\rightarrow D_{s}^{+}a_{0}^{-})$. In our QCDF calculation, where annihilation diagrams are power-suppressed and safely neglected, these two branching ratios are almost identical due to the intact $SU(3)_F$ flavor symmetry of the spectator quark. Conversely, in PQCD, the active W-exchange annihilation diagram in the $\overline{B}^{0}\rightarrow D^{+}a_{0}^{-}$ channel causes significant destructive interference, which is highly suppressed in the $\overline{B}_{s}^{0}\rightarrow D_{s}^{+}a_{0}^{-}$ mode due to flavor mismatch. This explains why the PQCD prediction for the $\overline{B}_{s}^{0}$ mode is prominently larger. Nevertheless, both approaches suffer from significant theoretical uncertainties arising from nonperturbative parameters, which can obscure the differences between the two scenarios. Therefore, more reliable nonperturbative methods and precise experimental data will be essential to unambiguously probe the nature of scalar mesons in future studies.
\end{itemize}
  \section{Summary}
  \label{sec04}
Inspired by recent advancements in heavy-flavor physics experiments and theoretical models, we have conducted a detailed analysis of the ${\overline B}_{d,s} \to D_{d,s}^{(*)+}S^{-}$~($S=K_0^*(1430), a_0(1450)$) weak decays. By incorporating QCD radiative corrections into the calculation of hadronic matrix elements within the QCDF framework, we systematically evaluated the decay amplitudes. Utilizing updated $B_{(s)} \to D^{(*)}$ form factors derived from the self-consistent CLF approach, our numerical results (summarized in Tables~\ref{tab:brS1} and \ref{tab:brS2}) indicate that most of these decay channels possess large branching ratios on the order of $\sim 10^{-5}$. Notably, the predictions for the ${\cal B}r({\overline B}^0 \to D^{+} a_0^{-})$ and ${\cal B}r({\overline B}_{s}^0 \to D_{s}^{+} a_0^{-})$ decays can reach as high as $\sim 10^{-4}$ in S2. Given these substantial decay rates, we strongly recommend that these specific channels be sought with high priority, as they stand out as prime candidates for early observation at the currently running LHCb and SuperKEKB experiments.
  \section*{Acknowledgments}

This work is supported by the National Natural Science Foundation of China (Grant No.12105078), the Natural Science Foundation of Henan Province (Grant No.262300421847) and the Doctoral Scientific Research Startup Fund of Anyang Normal University (Grant No.192096225010).

  \end{document}